\documentclass[aps,prc,showpacs,twocolumn,superscriptaddress]{revtex4}
\usepackage{graphicx,amsmath,amssymb}

\begin{document}


\title{Gauge Invariant Evaluation of Nuclear Polarization with
Collective Model}


\author{Yataro Horikawa}
\email[]{horikawa@sakura.juntendo.ac.jp}
\affiliation{Department of Physics, Juntendo University, 
Inba-gun, Chiba  270-1695, Japan }
\author{Akihiro Haga}
\email[]{haga@npl.kyy.nitech.ac.jp} 
\affiliation{Department of  Environmental Technology and Urban Planning, \\
Nagoya Institute of Technology,
Gokiso, Nagoya 466-8555, Japan}



\begin{abstract}
The nuclear-polarization (NP) energies 
with the collective model commonly employed in the NP calculations 
for hydrogenlike heavy ions are found to have serious gauge violations    
when the ladder and cross diagrams only are taken into account. 
Using the equivalence of charge-current density with a schematic 
microscopic model, the NP energy shifts with the collective model are gauge
invariantly evaluated for the $1s_{1/2}$ states in $^{208}_{~82}$Pb$^{81+}$ 
and $^{238}_{~92}$U$^{91+}$.  
\end{abstract}


\maketitle



High-precision Lamb-shift measurement on high-Z hydrogenlike atoms
\cite{BE95} spurred a renewed interest in the quantum electrodynamical
(QED) calculation of electronic atoms. Comparison of theoretical results 
with experimental data allows 
sensitive tests of QED in strong electromagnetic fields \cite{SA90,MO98}. 
In this context, the study of the nuclear-polarization (NP) effect
becomes important because the NP effect, as a non-QED effect which
depends on the model used to describe the nuclear dynamics, sets a limit 
to any high-precision test of QED. 

A relativistic field-theoretical treatment of NP calculation was presented 
by Plunien et al. \cite{PL89,PL95} utilizing the concept of effective 
photon propagators with nuclear-polarization insertions. In these
studies, only the Coulomb interaction was considered based on the
argument that the relative magnitude of transverse interaction is of the 
order of $ (v/c)^2$ and the velocity $v$ associated with nuclear
dynamics is mainly nonrelativistic. 

The effect of the transverse interaction was studied in the Feynman
gauge by Yamanaka et al.~\cite{YA01}
with the same collective model used in \cite{PL89,PL95,NE96} 
for nuclear excitations.
They found that the transverse contribution is several
times larger than the Coulomb contribution in heavy 
electronic atoms before the contributions of the 
positive and negative energy states cancel. 
However, due to the nearly complete cancellation between
them, the transverse effects become small and the net
effect is destructive
to the Coulomb contribution in both $1s_{1/2}$ states 
of $^{208}_{~82}$Pb$^{81+}$ and 
$^{238}_{~92}$U$^{91+}$.  
As a result, the total NP energy almost vanishes in 
$^{208}_{~82}$Pb$^{81+}$.

Recently, the NP effects for hydrogenlike  and
muonic $^{208}_{~82}$Pb$^{81+}$ were calculated in both
the Feynman and Coulomb gauges, using a microscopic random phase 
approximation (RPA) to describe nuclear excitations \cite{HHT02, HHT02a}.
It was found that, in the hydrogenlike atom, the NP effects due to 
the ladder and cross diagrams have serious gauge dependence and
inclusion of the seagull diagram is indispensable to restore the 
gauge invariance \cite{HHT02}. In contrast, 
the magnitude of the seagull collection is a 
few percent effect in the muonic atom,  
although it improves the gauge invariance \cite{HHT02a}.

In the present paper, 
we report that the nuclear collective model employed for hydrogenlike ions 
in \cite{NE96, YA01,PL89,PL95} 
also leads to a large violation of gauge 
invariance as far as the ladder and cross diagrams only are considered.
Then it is shown, based on the 
equivalence of the transition density 
of the collective model and 
a microscopic nuclear model with a schematic interaction 
between nucleons, that the seagull corrections should also be calculated
with the collective model in order to obtain gauge invariant NP results.
The resulting gauge invariant NP energy shifts are given  
for the $1s_{1/2}$ states in $^{208}_{~82}$Pb$^{81+}$ and $^{238}_{~92}$U$^{91+}$.

For spherical nuclei, the Hamiltonian of the small amplitude vibration 
with multipolarity $L$ is written 
as 
\begin{align}
 H_L = \frac{1}{2}
(\frac{1}{D_L} \sum_{M}\hat{\pi}_{LM}^\dagger  \hat{\pi}_{LM}
 + C_L \sum_{M}\hat{\alpha}_{LM}^\dagger \hat{\alpha}_{LM}),
\label{harmonic}
\end{align}
where $\hat{\pi}_{LM}$ are the canonically conjugate momenta to 
the collective coordinates $\hat{\alpha}_{LM}$.
The lowest vibrational modes are expected to have density variations with no radial
nodes, which may be referred to as shape oscillations.
The corresponding charge density 
operator with the multipolarity $L$ is written as 
\begin{align}
\hat{\rho}_L(t,\boldsymbol{r}) &= 
\rho_L(r) \sum_M Y_{LM}^* \hat{\alpha}_{LM}(t)
\label{rhoL}
\end{align}
to the lowest order of $\hat{\alpha}^\dagger _{LM}(t)$.

The liquid drop model of Bohr (BM)\cite{Bohr52} is a simple model of such a shape oscillation 
obtained by considering deformation of 
the nuclear radius parameter while leaving the surface diffuseness independent of angle:
\begin{align}
 R(\Omega) = R_0 \left[1 + \sum_{LM}  \alpha_{L M} Y^*_{LM}(\Omega)\right], 
\end{align}
where $R_0$ is the nuclear radius parameter of the ground state.
The radial charge density of BM
is given by   
\begin{align}
\rho_L(r) = - R_0 \frac{d }{dr}\varrho_0(\boldsymbol{r}),
\label{BMrho}
\end{align}
where $\varrho_0(\boldsymbol{r})$ is a charge distribution with spherical symmetry.

If we assume that under distortion, an element of mass moves from $\boldsymbol{r}_0$ 
to $\boldsymbol{r}$ without alteration of the volume it occupies, i.e., 
the nucleus is composed of an inhomogeneous incompressible 
fluid, a harmonic vibration of an originally spherical surface $r = r_0$ in the nucleus
is given by
\begin{align}
 r(\Omega) = r_0 \left[1 + \sum_{LM}  
\left(\frac{r_0}{R_0}\right)^{L-2} \alpha_{L M} Y^*_{LM}(\Omega) \right].
\end{align}
For this model we obtain 
\begin{align}
 \rho_L(r) = - \frac{1}{R_0^{L-2}}\  r^{L-1}\  \frac{d}{dr} \varrho_0(\boldsymbol{r}). 
\label{tassie}
\end{align}
This version will be hereafter referred to as the Tassie Model (TM) \cite{Tassie}. 
In Eqs.(\ref{BMrho}) and (\ref{tassie}), 
$\varrho_0(\boldsymbol{r})$  is usually taken 
to be equal to the ground-state charge distribution.

In either case, the motion of nuclear matter 
is assumed to be incompressible and irrotational,  
hence the velocity field $\boldsymbol{v}(t,\boldsymbol{r})$ is given by  
a velocity potential as
$\boldsymbol{v}(t, \boldsymbol{r}) = \boldsymbol{\nabla}\Phi(t,\boldsymbol{r})$.
This implies the nuclear current defined by 
$
\boldsymbol{J}(\boldsymbol{r})= \varrho_0(r) \boldsymbol{v}(\boldsymbol{r})
$
yields the transition multipole density of current operator 
\begin{align}
\hat{\boldsymbol{J}}_{L}(t,\boldsymbol{r}) &=  
J_{LL-1}(r)
\sum_M \boldsymbol{Y}^*_{LL-1M} \hat{\alpha}_{LM}(t).
\label{JL}
\end{align}
Note that the $J_{LL+1}(r)$ part  
does not appear in the transition density of current operator given by (\ref{JL}).

Therefore, in this kind of collective model, the continuity equation of
charge gives 
\begin{align}
[i \Delta E_L \rho_L(r)
 +  \sqrt{\frac{L}{2L+1}}(\frac{d}{dr}-\frac{L-1}{r})J_{LL-1}(r)] = 0, 
\label{continuity}
\end{align}
where $\Delta E_L $ is the excitation energy of the surface
oscillation. Hence the transition density of current is given by
\begin{align}
 J_{LL-1}(r) = i \Delta E_L \sqrt{\frac{2L+1}{L}} r^{L-1} 
\int_r^\infty x^{1-L} \rho_L(x) dx  
\label{current}
\end{align}
in terms of the transition density of charge.
If we assume the uniform charge distribution 
$\varrho_0(r) = \varrho_0\Theta(R_0 -r)$,
we obtain, for both BM and TM,  
\begin{align}
 \rho_L(r) &= \langle J_f\|r^LY_L\|J_i\rangle \  \frac{1}{R_0^{L+2}}\delta(R_0-r), 
\label{unitran}
\\
 J_{LL-1}(r) &= \langle J_f\|r^LY_L\|J_i\rangle \  i \Delta E_L \sqrt{\frac{2L+1}{L}} \frac{r^{L-1}}{R_0^{2L+1}}
\Theta(R_0-r).
\label{unicurrent}
\end{align}
The transition densities given by (\ref{unitran}) and (\ref{unicurrent})
have been employed in the previous NP
calculations for $L \ge 1$~\cite{PL89,PL95,NE96,YA01}. 
It should be mentioned that,  although the surface oscillation 
applies to the case of the multipolarity $L \ge 2$, 
Eqs.~(\ref{unitran})~and~(\ref{unicurrent}) with $L = 1$ give 
the transition densities of the giant dipole resonance 
given by the Goldhaber-Teller model 
describing the relative motion of neutrons 
and protons \cite{DW66}. 
For the monopole vibration, it is also possible to construct   
corresponding charge and current densities \cite{YA01, PL89}.

In general, the charge conservation relation between the charge and
current densities is necessary but not sufficient for the gauge 
invariance of the NP calculation. 
Unfortunately, it is practically impossible 
to construct a model that incorporates gauge invariance
explicitly in terms of the collective variables of the model. 
However, it is possible to evaluate the NP energy gauge invariantly   
with the above collective model as is shown below.

The NP calculations
with the collective model   
assume that a single giant resonance with spin multipolarity $L$
saturates the energy-weighted $B(EL)$ strength for each isospin.
In this respect, let us recall the fact that 
the transition densities of charge to the 
sum-rule saturated levels are given in terms of 
the ground-state charge density \cite{DF73}. This can be seen as follows.
For a pair of single-particle operators 
$g(\boldsymbol{r}) = g(r)Y_{LM}(\Omega)$ and $f(\boldsymbol{r}) =
f(r)Y_{LM}(\Omega)$, 
the energy-weighted sum rule can be generalized to 
\begin{align}
&\frac{1}{2J_i+1} \sum_n (E_n -
 E_i)[\langle J_n\|g(r)Y_L\|J_i\rangle ^*\langle J_n\|f(r)Y_L\|J_i\rangle ]
 \nonumber \\
&= \frac{2L+1}{4\pi}\frac{h^2}{2M}
\int r^2 dr \varrho_0(r)[g'(r)f'(r) + \frac{L(L+1)}{r^2}g(r)f(r)], 
\label{ewsumrule}
\end{align}   
where $\varrho_0(r)$ is the charge distribution of the ground state
normalized as $\int r^2 dr \varrho_0(r) = Z$ \cite{BM75}.
When a single excited state $ |J_fM_f \rangle$ saturates the $B(EL)$ strength, 
$ |J_fM_f \rangle  \propto r^L Y_{LM} |J_iM_i \rangle $, 
the transition density of charge to this state 
is derived from the sum-rule relation (\ref{ewsumrule}) model independently 
and given by   
\begin{align}
 \varrho_{fi}(r) = - \frac{1}{2L+1 }\ \frac{\langle J_f\|r^LY_L\|J_i\rangle }
{\langle J_i|r^{2L-2}|J_i \rangle } \ r^{L-1}\  \frac{d}{dr} \varrho_0(r).
\label{collectrho}
\end{align}
If the charge distribution of the ground state is assumed to be a uniform
distribution with a radius $R_0$, 
this becomes  
\begin{align}
\varrho_{fi}(r) = \langle J_f\|r^LY_L\|J_i\rangle \  \frac{1}{R_0^{L+2}}
\  \delta(r - R_0),
\end{align}
which is equal to the matrix element of the charge density 
operator of the collective model
given by (\ref{unitran}).

On the other hand, it is well known that 
the schematic RPA with a separable interaction 
\begin{align}
V_S(\boldsymbol{r}_i,\boldsymbol{r}_j) = \kappa_L 
\sum_M r_i^L Y_{LM}(\Omega_i)\ r_j^L Y^*_{LM}(\Omega_j).
\end{align}
for particle-hole excitations $|m i^{-1}\rangle$ with a degenerate particle-hole excitation energy $\epsilon$
gives a collective state 
$|LM \rangle$, 
which exhausts the energy-weighted sum rule for the single particle operator $r^L Y_{LM}$:
\begin{align}
\Delta E_L\ |\langle LM|r^L Y_{LM}|0\rangle |^2 = \epsilon \sum_{mi} 
|\langle m|r^L Y_{LM}|i\rangle |^2.  
\end{align}
where  $\Delta E_L$ is the excitation energy of $|LM \rangle$~\cite{RS80} . 
If the ground state is assumed to be 
a filled major shell of the 
harmonic oscillator potential:
\begin{align}
 H_{HO} = \frac{1}{2M_N}\boldsymbol{p}^2 + \frac{M_N\omega^2}{2} \boldsymbol{r}^2,
\end{align}
the particle-hole excitation energy $\epsilon$
is taken to be $1 {\hbar}  \omega$ for $1^-$ 
and $2 \hbar \omega$ for $0^+$ and $2^+$. 
The corresponding collective states exhaust the energy-weighted sum rules,
because the transition strengths 
vanish outside these p-h excitation spaces.  
Therefore, the transition densities of charge to the collective
states of this fictitious nucleus are given by (\ref{collectrho}). 
When the ground-state charge density is approximated 
by a uniform charge density, 
the transition density of charge becomes identical with
that of the collective model employed in NP calculations for
hydrogenlike atoms.
However, the gauge invariant electromagnetic interaction of 
this schematic microscopic model is given by the minimal substitution 
$\boldsymbol{p}_i \rightarrow \boldsymbol{p}_i - e_i \boldsymbol{A}$
to the Hamiltonian 
$  H = H_{HO} + V_S $.
Hence the lowest-order contributions to NP with this model  
are given by the three Feynman diagrams in Fig.~1, where 
two photons are exchanged between a bound electron and a nucleus.
The nuclear vertices are understood to have no 
diagonal matrix elements for the ladder and cross diagrams,
and no nuclear intermediate states for the seagull diagram.
It is well known that the NP results with this model is gauge 
invariant provided these three diagrams are taken into account. 
Although $J_{LL+1}(r)$ current density appears in this model,
$J_{LL-1}(r)$ dominates in the transition to the collective state.

Thus we can conclude that the gauge invariance of the collective
model is also guaranteed with the charge-current density satisfying 
the continuity equation~(\ref{continuity}), 
provided the contributions from the three
diagrams are taken into account.   
It should be noted that the seagull contribution is given 
in terms of the ground-state charge distribution 
and does not depend on the details of the model for nuclear excitations.
In the actual NP calculations \cite{NE96,PL89,PL95,YA01} with the collective
model, the assumption that each nuclear intermediate state 
saturates the sum-rule is not strictly obeyed, 
because the observed nuclear data are used for the low-lying states. 
However, since the gauge violation is serious only in the dipole 
giant resonance, this does not invalidate our arguments as is confirmed
by the numerical results in the following.
\begin{figure}
  \includegraphics[width=7cm]{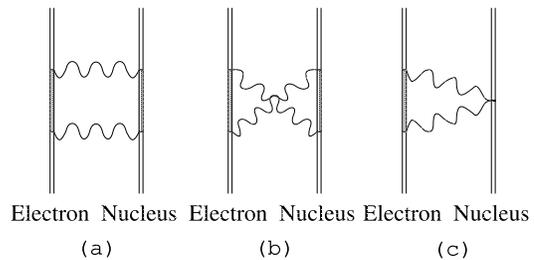}
  \caption{Diagrams contributing to nuclear polarization in lowest order; (a) ladder, (b) cross and (c) seagull diagrams.}
  \label{fig;npdiagram}
\end{figure}

The formulas to calculate the NP energy shifts due to the three 
diagrams of Fig.~1 were given in \cite{HHT02} for arbitrary nuclear models.
In the present NP calculations of the  
$1s_{1/2}$ states in hydrogenlike $^{208}_{~82}$Pb and 
$^{238}_{~92}$U,  
the parameters of the collective model are the same as
those given in Refs. \cite{NE96,YA01}. 
The same low-lying states and giant resonances are taken into account.
In addition, the contributions from the 
$4^-$ and $5^-$ giant resonances are also calculated in order 
to see the effects of higher multipoles neglected previously.
The $B(EL)$ values are adjusted to the observed values for low-lying states and
the $B(EL)$ are estimated through the energy-weighted 
sum rule for giant resonances. 

Tables I and II show the results for the $1s_{1/2}$ states in  
$^{208}_{~82}$Pb$^{81+}$
and   $^{238}_{~92}$U$^{91+}$, 
where the sum of the 
contributions from the three diagrams of Fig.1 
is given for each multipole.
The second and the third columns are the results 
including the transverse effects in the Feynman and Coulomb 
gauges, respectively.
The values in the parentheses are the contributions
from the seagull diagram. 
The NP energy shifts due to the ladder and crossed diagrams only 
are obtained by subtraction of the seagull contributions given
in the parentheses.
The fourth column gives the results of the 
present Coulomb nuclear polarization (CNP).
The last two columns are the results of the previous
calculations. 

\begin{table}
\caption {Nuclear-polarization correction (meV) to the $1s_{1/2}$ state 
 of $^{208}_{~82}$Pb$^{81+}$. NP denotes the correction due to the whole
 of the Coulomb and transverse  interactions; CNP the correction only
 due  to the Coulomb interaction. Energy shifts in the parentheses 
are due to seagull contribution.}
\begin{ruledtabular}
\begin{tabular}{crrrrrrr}
& \multicolumn{5}{c}{present~work}&
\multicolumn{1}{c}{Ref.~\cite{YA01}}&
\multicolumn{1}{c}{Ref.~\cite{NE96}}
\\
\multicolumn{1}{c}{$L^{\pi}$}&
\multicolumn{2}{c}{Feynman(NP)}&
\multicolumn{2}{c}{Coulomb(NP)}&
\multicolumn{1}{c}{CNP}&
\multicolumn{1}{c}{NP}&
\multicolumn{1}{c}{CNP}\\
\hline
$0^+$
&-3.3&(-0.2)~~
& -3.3&(+0.0)~~
& -3.3 
& -6.6 & -3.3
\\

$1^-$ 
& -22.1&(-42.3)~~
& -21.5&(-7.3)~~
& -17.0
&+16.3 & -17.6 
\\

$2^+$ 
& -5.8&(+0.3)~~
& -5.8&(+0.6)~~
& -5.8 
&-7.0 & -5.8
\\

$3^-$ 
& -2.7&(+0.2)~~
& -2.8&(+0.2)~~
& -2.9
&-2.9&-2.6 \\

$4^+$ 
& -1.0&(+0.1)~~
& -1.0&(+0.1)~~
& -1.1
&& 
\\

$5^-$ 
& -0.5&(+0.1)~~
& -0.6&(+0.0)~~
& -0.6
&& \\
\hline 
{\rm total}
& -35.4&(-41.8)~~
& -35.0&(-6.4)~~
& -30.7
& -0.2 & -29.3 
\end{tabular}
\end{ruledtabular}
\end{table}

\begin{table}
\caption {Nuclear-polarization correction (meV) to the 
$1s_{1/2}$ state of $^{238}_{~92}$U$^{91+}$. 
The notations are the same as in Table I.}
\begin{ruledtabular}
\begin{tabular}{crrrrrrr}
& \multicolumn{5}{c}{present~work}&
\multicolumn{1}{c}{Ref.~\cite{YA01}}&
\multicolumn{1}{c}{Ref.~\cite{NE96}}
\\
\multicolumn{1}{c}{$L^{\pi}$}&
\multicolumn{2}{c}{Feynman(NP)}&
\multicolumn{2}{c}{Coulomb(NP)}&
\multicolumn{1}{c}{CNP}&
\multicolumn{1}{c}{NP}&
\multicolumn{1}{c}{CNP}\\
\hline
$0^+$ 

& -9.3&(-0.4)~~
& -9.3&(+0.0)~~
& -9.3 
& -21.5 & -9.5
\\

$1^-$ 
& -54.3&(-65.7)~~
& -52.5&(-3.9)~~
& -41.6
&-3.8 & -42.4 
\\

$2^+$ 

& -131.6&(+0.0)~~
& -131.7&(+1.6)~~
& -131.6 
&-148.2 & -138.9
\\
$3^-$ 
& -6.5&(+0.3)~~
& -6.5&(+0.4)~~
& -6.7
&-7.3&-6.8 \\

$4^+$ 
& -2.0&(+0.2)~~
& -2.0&(+0.2)~~
& -2.1
& & \\

$5^-$ 
& -1.0&(+0.1)~~
& -1.0&(+0.1)~~
& -1.1
& & \\
\hline 
{\rm total}
& -204.7&(-65.5)~~
& -203.0&(-1.6)~~
& -192.4
& -180.8 & -197.6 
\end{tabular}
\end{ruledtabular}
\end{table}

The results  with the collective model, 
as with the microscopic RPA model \cite{HHT02},  
also lead to large violations of gauge invariance 
if ladder and crossed diagram contributions only are
considered. The seagull corrections are considerable 
in the $1^-$ contributions for both of $^{208}_{~82}$Pb$^{81+}$ and 
$^{238}_{~92}$U$^{91+}$.
Note that, in the limit of point nucleus, 
which is not unrealistic even for heavy
hydrogenlike ions, the seagull collection occurs only 
in the dipole mode which involves the current density $J_{10}(r)$.

In $^{208}_{~82}$Pb$^{81+}$, 
the contributions from low-lying states are about 10\% 
of the total results
and the NP energy shift is mainly 
determined by the giant resonance contributions.
The most dominant contribution comes 
from the giant dipole resonance, 
where a large violation of gauge invariance 
occurs if the seagull contributions in the parentheses are neglected:
$- 22$ meV becomes $+ 20$ meV and $- 14$ meV
in Feynman and Coulomb gauges, respectively.
The column 5 gives the previous results in the Feynman
gauge without seagull contributions. 
The differences between the two results in the Feynman 
gauge without seagull contribution 
come from the accuracy of numerical integration over 
the continuum threshold region of 
electron intermediate states 
and from the differences of the electron wave functions: here we have used
wave functions in a finite charge distribution,  while \cite{YA01} 
employs point Coulomb solutions. 

In $^{238}_{~92}$U$^{91+}$, 
the dominant contribution comes from the lowest excited states
$2^+$ with a large $B(E2)$ value. 
Since the transition density of current in the present model given by (\ref{unicurrent}) is proportional to the excitation energy, 
the transverse contribution of the lowest $2^+$ is negligible 
due to its exceptionally small excitation energy 
$\Delta E_2 = 44.9$ keV.
Apart from this large Coulomb contribution, 
the contributions from other states show similar tendencies as
in $^{208}_{~82}$Pb$^{81+}$.
Namely, the contributions from other low-lying states 
are small compared with the giant resonance contributions, 
and a large gauge violation occurs in the giant dipole resonance
when the seagull contribution is omitted.

To summarize, the transverse effects with the collective model
are estimated gauge invariantly by inclusion of the seagull contribution. 
The gauge invariance is satisfied    
to a few percent levels in both 
$^{208}_{~82}$Pb$^{81+}$ and $^{238}_{~92}$U$^{91+}$ for 
each of the multipoles separately.
Without the seagull correction, 
the Feynman gauge in particular does not give reliable 
predictions of NP, although numerical calculation in this gauge is 
easier than in the Coulomb gauge.
Hence the conclusion of \cite{YA01} on the transverse effects is no longer tenable.  
The NP energy shifts are $-35.0 (-35.4)$ meV in 
$^{208}_{~82}$Pb$^{81+}$ and $-205 (-203)$ meV in
$^{238}_{~92}$U$^{91+}$ for Coulomb (Feynman) gauge.
The net transverse effect is about $14 \sim 15 \%$ of the Coulomb
energy shift of $-30.7$ meV in $^{208}_{~82}$Pb$^{81+}$. 
This is similar to the conclusion of the
microscopic model \cite{HHT02}, and should be compared with the 
transverse effect of the $1s_{1/2}$ 
state in muonic $^{208}_{~82}$Pb, which is about 
6 \% of the Coulomb contribution \cite{HHT02a}.
The agreement between the two models provides stability of the 
prediction of the NP effects with respect to the choice of the nuclear models.
The percentage of the transverse effect 
in the total shift in $^{238}_{~92}$U$^{91+}$ is reduced to about 
$6$\% of the Coulomb effect due to the dominant 
Coulomb contribution from the lowest $2^+$ state.

The authors wish to acknowledge Prof. Y. Tanaka for generous support 
and useful discussions during our research on NP effects.  
They appreciate Drs. N. Yamanaka and A. Ichimura 
for collaboration on the NP effects with the collective model, 
which motivated the present work.

\end{document}